# Localized all-optical control of single semiconductor quantum dots through plasmon-polariton-induced screening

*Matt Seaton[1], Alex Krasnok[2], Allan S. Bracker[3], Andrea Alù[2], and Yanwen Wu[1*]*

1. Department of Physics and Astronomy, University of South Carolina, Columbia, SC 29208

2. Department of Electric and Computer Engineering, University of Texas at Austin, Austin, TX 78712

3. Naval Research Laboratory, Washington DC 20375-534778712

* Email: wu223@mailbox.sc.edu

*Due to their ability to strongly modify the local electromagnetic (EM) field through the excitation of surface plasmon polaritons (SPPs), plasmonic nanostructures have been often used to reshape the emission direction and enhance the radiative decay rate of quantum emitters, such as semiconductor quantum dots (QDs). These features are essential for quantum information processing, nanoscale photonic circuitry and optoelectronics. However, the modification and enhancement demonstrated thus far often drastically alter the local energy density of the emitters, and hence their intrinsic properties, leaving little room for active control. Here, we demonstrate dynamic tuning of the energy states of a single semiconductor QD by optically modifying its local dielectric environment with a nearby plasmonic structure, instead of directly coupling it to the QD. This method leaves the original intrinsic optical properties of the QD intact, enabling the opportunity of tuning its optical properties in real time. This capability is highly desired in applications requiring ultrafast switching and modulation mechanisms.*

Single quantum emitters such as semiconductor QDs provide a robust and stable physical confinement for real (electrons) and quasi (excitons, holes) particles. Transitions between electronic states, discretized by the confinement, lie in the near-infrared (NIR) and visible wavelength range, making it possible for repeatable ultrafast optical excitation and coherent manipulation within a given dot. These properties make QDs attractive candidates in applications such as on-demand single photon generation[1–5] and physical implementation of qubit operations for quantum computation[6–13]. However, in applications where active control is required, the ability to tune in real time the linear and nonlinear optical properties is essential. For instance, modifying the QD energy states through its confinement potential is a key ingredient in the generation of entangled photon pairs[4,14] and for the creation of entanglement in QD-based quantum information processing[12]. This flexibility is difficult to achieve in QDs grown using self-assembled processes, such as molecular beam epitaxy (MBE) or chemical synthesis. Once the QDs are grown, their confinement potentials are determined by their post-growth physical structures and dielectric compositions, making it difficult to effectively change the energy levels within the dot or tune the coupling between spatially separated dots. One way to address this challenge is to define the lateral dot confinement potentials post-growth through lithographically fabricated electrical gates on quantum wells[15–17]. By electrically tuning the gate voltage, the confinement potential can be dynamically modified to tune the energy of the gate-defined dot, as well as switch between different charged states. However, the disadvantage associated with this system is that the gate-defined confinement potential is controlled by electronics which limits the control speed. Furthermore, most gate-defined dots do not support excitons, and hence makes it difficult to perform optical manipulation. An ideal system would possess an inherently robust and stable confinement potential when tunability is not required. When tunability is desired, the system should be able to switch at an ultrafast optical frequency to a different confinement state. Such a solution can be achieved through the use of plasmonic structures, as we discuss in the following.

Plasmonic structures have been shown to enhance electromagnetic (EM) fields in their vicinity through the excitation of surface plasmon polaritons (SPPs) at a metallic-dielectric interface. When a QD is brought near a plasmonic surface, the change in its local optical density of states can drastically affect and enhance its radiative and nonradiative lifetimes[18–26]. Consequently, these enhanced decay mechanisms tend to dominate the emission dynamics of the QDs, and often obscure their intrinsic optical properties. In addition, the environment around the QD considered following this approach has been inherently static so far. Therefore, while the environment plays an important role in defining the properties of the SPP field that modifies the QD, its contribution is not easily tunable. If this dielectric environment can be dynamically and optically manipulated as well, then its active role can lead to interesting consequences. Here, we demonstrate an approach to locally manipulate the dielectric environment of single QDs by optically exciting a plasmonic structure embedded in the MBE grown heterostructure. In this case, the SPP enhancement field does not affect the QDs directly, because the emitter is located just beyond the SPP near field decay length[27]. Rather, the plasmonic structure is excited to induce changes in the dielectric surroundings of the single QDs, which in turn tunes the optical properties of the QDs. We report that this tuning effect is localized near the optically excited plasmonic interface, which is smaller than the spot size of the laser. Essentially, the optically excited plasmonic structure mimics the effect of a nanoscale electrical gate. This introduces a new tool to realize dynamic optical manipulation of a single QD for applications in optoelectronics, photonics, and information processing.

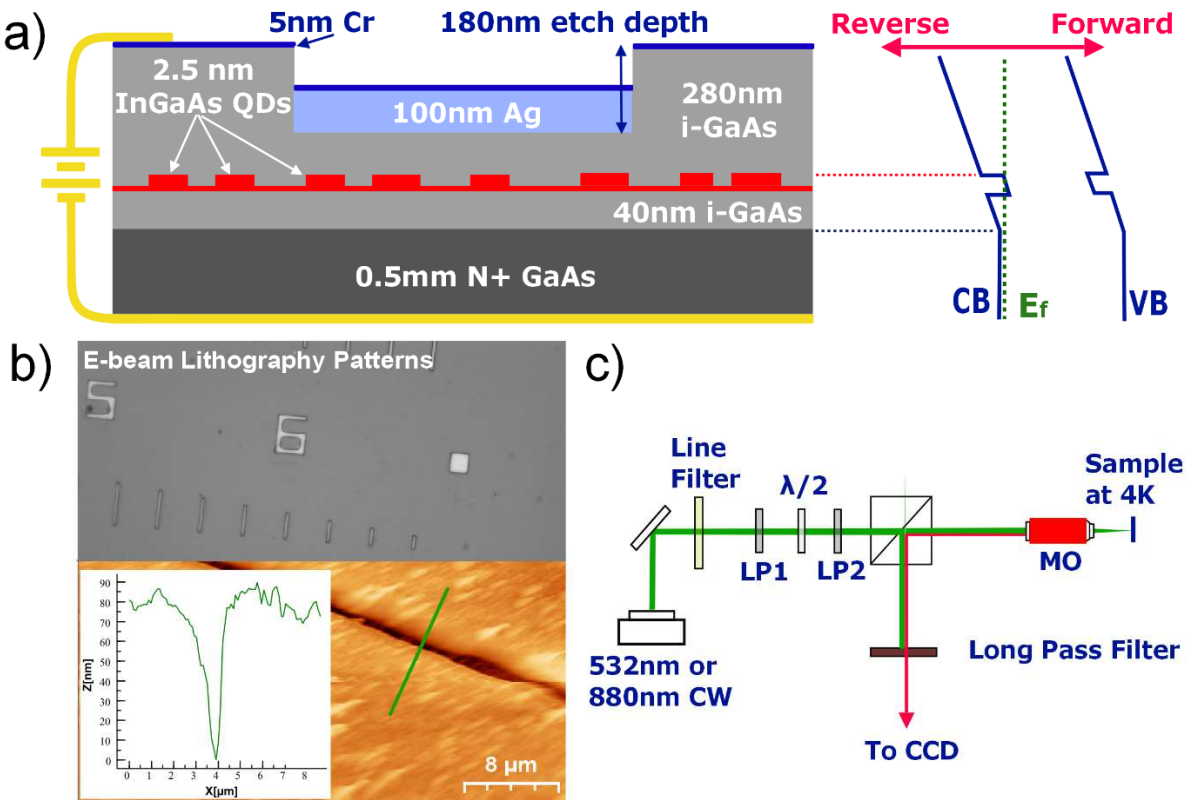

Figure 1: (a) Post fabrication structure of the self-assembled QD sample and the corresponding energy band diagram of the Schottky diode. CB (VB) is the conduction (valance) band and $E_f$ marks the Fermi energy. (b) Optical micrograph, AFM scan, and topography slice of the Ag filled groove, (c) Confocal PL experimental setup. LP is linear polarizer and MO is the 50x microscope objective.

**Results**

The sample used in our experiment consists of MBE grown disk-shaped InGaAs QDs inside a Schottky diode structure with silver (Ag) slabs embedded in the capping layer. The detailed sample fabrication process is described in the Method section. The fabricated sample structure and the Schottky diode schematics are shown in Figure 1a. An atomic force microscopic (AFM) scan and topography measurement of the sample surface shown in Figure 1b confirm the etched groove depth and deposited Ag thickness. We performed polarization, power and bias voltage dependent studies on QDs on and away from the embedded Ag slabs through a confocal setup illustrated in Figure 1c. Since a slab of 100 nm of Ag is not transparent, the laser field does not penetrate the slab for on-slab excitation and hence only QDs near and to the side of the Ag slab can be excited and observed, as shown in Figure 1a. The Ag slabs are also electronically isolated from the Schottky contact. Detailed description of this setup can be found in the Methods section. When a bias voltage is applied across the sample through the deposited Schottky diode, we are able to charge the single dots in a controlled fashion with the desired number of electrons or holes. The initial charging of the dot ground state determines the final exciton species optically detected in our PL measurements[28–30]. For example, an uncharged ground state leads to a neutral exciton, an electron and hole pair, whereas a ground state charged with a single electron (hole) leads to a negative (positive) trion with two electrons (holes) and one hole (electron). The different exciton species are energetically distinct from each other. When the sample is forward biased, additional electrons are transferred into the QD from the Fermi sea, while reverse bias removes electrons. We monitor changes in these charging states within the Schottky diode structure to infer information on the local field environment.

Figure 2a shows a typical comparison of PL spectra obtained using both vertically (TE) and horizontally (TM) polarized excitations directly on and away from an Ag slab using 532 nm excitation. The TE (TM) mode is defined to be polarized parallel (perpendicular) to the edge of the Ag slab. The spectra collected away from the Ag slab show the spectral signature of a quasi-continuum of states instead of clear emission peaks from single dots. When the QDs are excited near the slab, however, there is a significant reduction in the quasi-continuum background and the single dot emission peaks became visible. This drastic difference in the spectra can be mainly attributed to screening by the excessive carriers generated in the thick GaAs capping layer. The removal of 140 nm of the capping layer above the Ag slab reduces this effect. We note that the field enhancement effect from the plasmonic Ag slab is unlikely to play a direct role in the increased single dot signal, due to its relatively large spatial separation (100 nm) from the plane of the dots[27].

A notable feature in the on-slab excitation spectra in Figure 2a is the significant polarization dependence of some of the single dot emission peaks, as indicated by the purple arrows. This behavior is not expected in highly symmetric disk-shaped QDs, where any fine structure splitting in the radiative spectrum due to strain is far less than the resolution of our optical detection setup[29]. To confirm this, we performed PL measurements on a similar sample with both empty and Ag-filled grooves and did not observe similar polarization dependent features from the empty grooves (see <mark>Supplemental Information</mark> for details). Therefore, the only symmetry breaking factor here is the embedded Ag slab.

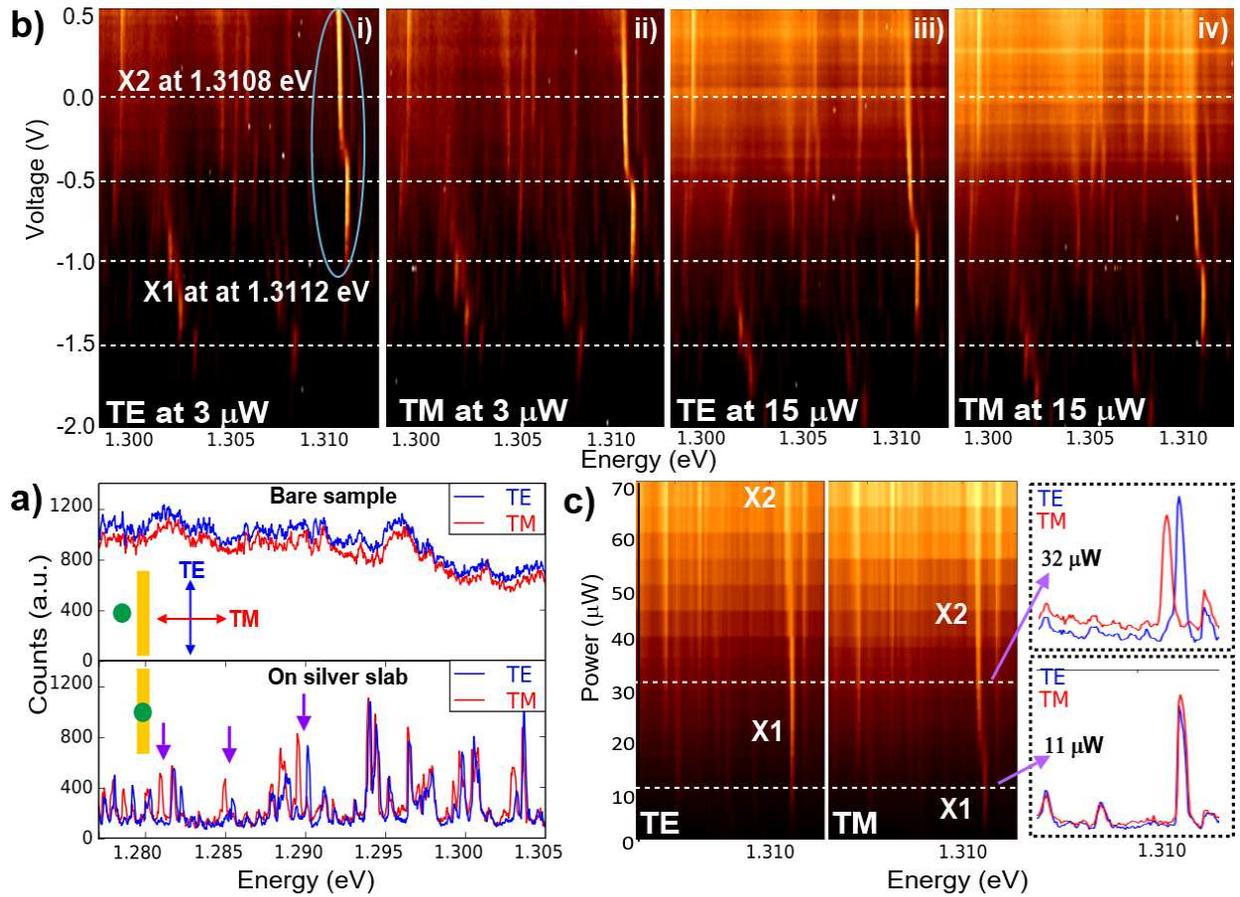

**Figure 2** Voltage and power dependent PL spectra from the hybrid structure. a) PL taken on and off the Ag slab at -1.4V bias with 8 μW excitation. b) Voltage dependent PL maps at TE and TM polarization with two different excitation power. c) Power dependent PL maps taken at -1V bias. PL from a) is taken from a different Ag slab.

In order to further investigate the effect of the Ag slab, we performed a series of bias voltage and power dependent PL experiments for both TE and TM excitation polarizations on and near an Ag slab. First, we used two excitation power levels, at 3 μW and 15 μW, as measured before the microscope objective, to generate bias-dependent PL maps for both TE and TM polarizations, as shown in <mark>Figure 2b.</mark> The evolution of the spectral lines reflects how different exciton species change with respect to the applied bias voltage. The most prominent feature in each map is the shift of an emission line from <mark>1.3108 eV, labeled by X1, to 1.3112 eV,</mark> labeled by X2, as indicated in the blue oval. This voltage-dependent change in emission energy corresponds to switching between two different exciton species within a single QD. The exact species are difficult to discern without more detailed studies. However, their identification is irrelevant in this case. Our focus, rather, is on the relative voltage at which the transition occurs under different excitation conditions. There are two notable trends visible in the four voltage dependent PL maps in <mark>Figures 2b</mark>. First, as the laser power is increased from 3 μW to 15 μW, the transition between X1 and X2 shifts to a more negative voltage for both polarizations. For TE polarization, this transition occurs <mark>at -0.3 V</mark> for 3 μW, as seen in Figures 2bi, and at <mark>-0.75 V for</mark> 15 μW, as seen in Figures 2biii. Similarly, the same transition shifts from -0.5 V to -1.0 V for TM polarization, as seen in Figures 2bii and 2biv, respectively. In general, all emission lines experience a negative trending shift as excitation power increases. Second, by maintaining the incident power and varying the polarization from TE to TM, a similar negative shift is observed in both excitation powers for the X1 and X2 transition. This can be seen by comparing Figures 2bi to 2bii and Figures 2biii and 2biv. More interestingly, in contrast to the power dependent case, we note that not all emission lines exhibit a polarization dependence shift. In particular, the unaffected states typically show weaker emission signals.

Next, we obtained the power dependent PL maps for the most prominent feature in the emission spectra, the transition from X1 to X2, for both TE and TM excitation polarization at a fixed bias of -1 V. The results presented in Figure 2c clearly show that the evolution between the two states as a function of power are starkly different for the two polarizations. Specifically, a complete transition from X1 to X2 occurs at the excitation power of 20 µW for TM polarization, which is roughly three times lower in power compared to the nearly 70 µW for TE. Furthermore, the completion of this transition requires a much smaller change in power for TM (< 5µW) compared to TE (>50µW). Single spectra are selected at excitation powers of 11 µW and 32 µW from each polarization and compared in Figure 2c. At 11 µW, the state emits at X1 for both TE and TM polarizations. At 32 µW, however, the transition from X1 to X2 is completed for TM but has not yet begun for TE. This polarization dependence at different excitation powers is responsible for the polarization dependent features seen in Figure 2a.

## Discussion

Inspecting the observed power and polarization dependent effects, the former has a straightforward explanation. With above-band 532 nm excitation, the number of free carriers increases linearly with excitation power. These free carriers rearrange spatially to screen the electric field in the Schottky structure. This screening leads to the flattening of the GaAs band and manifests itself as a general negative shift of the emission lines in the voltage map. As the carrier density reaches a certain threshold where the external field is completely screened, this effect and the shift towards negative voltages eventually saturates as a function of power. This is evident in the TE power dependent PL map of Figure 2c where the transition from X1 to X2 requires a large change in power of at least 50 µW.

While the power dependent screening effect influences all QDs under the optical excitation, the polarization dependent effect, on the other hand, only affects a few QDs, as seen in both Figure 2a and 2b. This implies that polarization selectivity and spatial selectivity are inherently connected. First, we eliminate the most obvious explanation, which is that the polarization dependence is simply due to the difference in the diffracted field intensity at the dot plane under TE and TM excitation. While this is indeed the case in the simulations with straight groove edges, we find that this difference is reduced when the groove edges are rounded in a more realistic model of our nanostructures, which have been realized using the wet-etch technique (See Supplemental Information for comparison). Even if the polarization difference is indeed a contributing factor in the simulation, it does not explain the dominant response observed in the experiment, since the simulation and experiment show opposite trends when the polarization is reversed.

A more fitting explanation that addresses the drastic polarization dependence and the spatial selectivity is related to the plasmonic effect from the edge of the Ag slab. Since SPPs are longitudinal modes, due to energy and momentum conservations, they can only be excited with TM waves. Given the normal incident geometry of the laser, only TM polarized light is capable of launching SPPs down the vertical edges of the Ag slab. As seen in the simulations in Figure 3a, the launched SPPs are confined to the Ag/GaAs interface. Hence, any plasmonic related effect is only experienced by QDs in the vicinity of the illuminated site of the interface. This explains why the fainter emissions in Figure 2b, likely from QDs located farther from the Ag slab and the center of the excitation spot, exhibit evidence of power dependent shifts but show no signs of polarization dependence. The polarization dependent negative shift of only selected QDs in the PL maps strongly suggests the addition of a *localized* screening of the applied electric field upon the optical launching of SPPs down the vertical edge of the Ag slab under TM excitation.

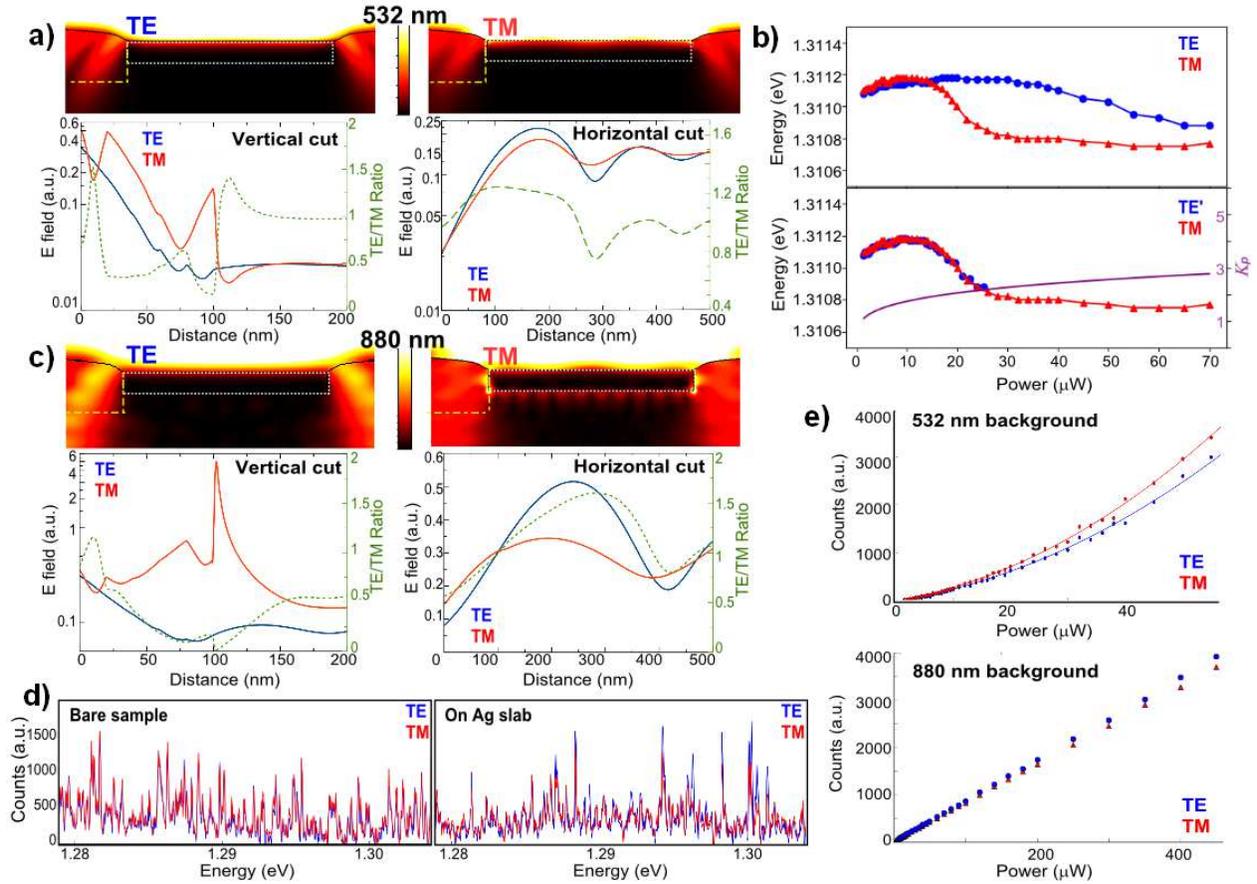

**Figure 3** a) and c) FDTD simulations of both TE and TM excitations at 532 nm and 880 nm. The dotted boxes indicate the location of the silver slab. The 1D plots are taken from the dashed vertical (along with silver interface) and horizontal (at dot layer) lines in the corresponding 2D simulations above. b) Top: Peak energy vs excitation power for both TE and TM excitations extracted from Figure 2c. Bottom: The same plot with the TE power rescaled with scaling factor $K_p$ to best overlap the TE and TM plots. d) PL spectra taken with 500 µW of 880 nm excitation at -1 V bias on the bare sample and on the Ag slab. e) Quasi-continuum backgrounds signals vs. excitation powers for 532 nm and 880 nm excitations.

This screening is the result of the extra free carriers generated by the SPP field in the GaAs capping layer. In general, emitters coupled to SPPs in a plasmonic structure experience enhancement in both excitation and recombination efficiencies[20,21,31]. With 532 nm above-gap excitation, TM polarization simultaneously generates SPPs on the Ag/GaAs interface and free carriers in the GaAs capping layer whereas TE polarization only generates the free carriers. The SPPs produce hotspots of enhanced EM field concentrated near the Ag slab as seen in Figure 3a. This field increases the local carrier density, which in turn is responsible for the additional localized screening experienced by nearby QDs. In Figure 3b, we plot the peak energies of the X1 and X2 transitions from Figure 2c as a function of power. The same graph is replotted by rescaling the power of the TE data using an empirical function of $P'_{TE} = P_{TE}/K_P$, where $K_P = (P_{TE})^{0.24}$, to provide the best overlap for TE and TM excitations. The scaling factor $K_P$ can be considered as the equivalent amount of extra power needed for TM light inside the sample to produce the same shift as TE. We see in Figure 3b that $K_P$ is power dependent and its value saturates around 3 at high power. We also extracted the quasi-continuum background as a function of power in Figure 3d for both polarizations. These background curves are good indications of the relative field strengths of the two polarizations inside the sample. They are indeed comparable to the simulations and do not differ by the scaling factor $K_P$. This is further proof that the observed behavior presented in Figure 2 is not merely a result of more light coupling to the sample for TM excitation, but rather an effect of extra SPP-induced screening. This SPP-induced local screening has multiple advantages over the power dependent screening effect from the perspective of performing optical control on the electrical environment of a QD. First, the localized screening effect is concentrated to the vicinity of the SPP, which can be controlled on the nanoscale by the shape of the plasmonic structure. Second, the localized screening effect requires less excitation power. We considered other possibilities involving non-SPP related interactions between the metal and free carriers in the GaAs, such as charge trapping and diffusion at a Schottky interface[32] and an electrostatic interaction with image charges in the metal[33]. However, none of these effects is expected to exhibit polarization dependence.

To truly pinpoint the role of free carriers, we performed PL studies of 880 nm below-band excitations to suppress free carriers in the GaAs capping layer. As expected, we found that, without free carriers in the GaAs, the signature polarization dependent screening behavior under 532 nm excitation is no longer observed. However, we note that the PL spectra under 880 nm excitation are not completely polarization independent. The emission peaks under TM excitation on the Ag slab on average have lower intensity counts as seen in <mark>Figure 3c</mark>. In contrast, PL spectra from the bare sample are completely polarization independent, as expected based on the cylindrical symmetry of the QDs. Statistics from a collection of below-band PL maps from different sites on and away from the Ag slabs show that this trend is general (see <mark>Supplemental Information</mark>). The exact cause for the reduced emission in the TM spectra is unclear. At first glance, the simulations suggest that this is simply the effect of the difference in diffracted field strength for TE and TM polarizations at the dot plane, as shown in the simulation in <mark>Figure 3c</mark>. However, if that were indeed the case, then this polarization intensity difference should also be reflected in the background. On the contrary, the quasi-continuum background due to the wetting layer for both TE and TM 880 nm excitation shows little to no difference in <mark>Figure 3d</mark>. In fact, the 880 nm PL data prior to background subtraction further confirm this discrepancy (see <mark>Supplemental Information</mark>). Therefore, additional dynamics are responsible for the overall peak intensity difference. From the simulations in <mark>Figure 3c</mark>, we see that the SPP field for 880 nm is much stronger than 532 nm. A possible explanation is the coherent coupling of the SPP field gradient and the mesoscopic moment (higher order poles) of the QD[27]. This pathway can interfere either constructively or destructively with the dipole coupling term, which leads to either enhancement or suppression of the radiative rate of the QD. In our dot/metal orientation, this interaction is destructive, hence leading to an overall decrease in emission. However, due to the fact that the silver slab is far from the QD where the SPP field is vanishing, further experiments with closer silver slabs are needed to quantify this effect. The ability to manipulate the SPP gradient field on the nanoscale will provide unprecedented optical probing and control to the elusive magnetic-dipole[34] and forbidden higher order transitions[35] in a quantum emitter.

## Conclusion

In summary, we have demonstrated that plasmonic nanostructures embedded in an optically controllable dielectric environment can be used to dynamically control QD emission. Our study suggests that the area affected is smaller than the optical spot size and it is localized near the launch site of the SPP. In addition, with the introduction of Ag slabs, the SPP-induced screening effect requires nearly three times less excitation power than power tuning alone. However, this factor is by no means the limit, and our results represent a proof-of-principle demonstration. This effect can be improved further by optimizing the design of metallic structures to shape and control the nanoscale electrical landscape around emitters. This tuning method, which is completely optical, does not suffer the various drawbacks of electrical control, specifically complicated nanofabrication of large numbers of electrical components, slow electrical switching speed, electrical noise, etc. Furthermore, unlike other quantum emitters and plasmonic hybrid structures, the Ohmic losses suffered by the plasmonic structure in this configuration do not obscure the intrinsic properties of the quantum emitter. Due to the complex many-body interactions possible in this hybrid system, further studies are required to quantify all the individual microscopic mechanisms giving rise to the total SPP-induced effect. Yet, the outcome of the effect is clear in our measurements, opening a promising pathway towards plasmonic-assisted low power and dynamical optical switching applications in the fields of optoelectronics, photonic circuitry and quantum information processing.

## Method

**<u>Sample fabrication:</u>** The sample layer structure was grown on an n-typed GaAs substrate and consists of a 500 nm Si-doped (n-type) GaAs buffer, 40nm of undoped GaAs, InGaAs QDs, and a 280 nm undoped GaAs capping layer as illustrated in Figure 1a. The InGaAs QDs are nominally disk-shaped with diameter of 10-20 nm and height of 2.5 nm. Rectangular silver (Ag) plasmonic slabs 1 μm wide, 100 nm thick, and various lengths (5 – 50 μm) were embedded in the GaAs capping layer through a combination of nanofabrication techniques. First, we coated the sample with a 200 nm thick layer of polymethyl methacrylate (PMMA). Then, E-beam lithography was used to write the patterns on the PMMA. The pattern was developed by a 3:1 mixture of isopropanol (IPA):methyl isobutyl ketone (MIBK) for 30s. Next, we employed a weak phosphoric acid and peroxide wet etch solution ($H_3PO_4$:$H_2O_2$:$H_2O$ = 1:4:50) to etch the exposed patterns into the GaAs capping layer. The final etched depth of 180 nm was achieved at an etch rate of 5nm/s. This depth was strategically chosen to place the plasmonic structure just outside of the range of the plasmonic field penetration depth (~100 nm)[27]. We then deposit the Ag slabs in the etched grooves using thermal vapor deposition. After liftoff of the PMMA, a 4nm semi-transparent chromium layer was deposited on the capping layer to form the top Schottky contact. An Ohmic contact was made on the doped substrate at the underside of the sample. A bias voltage can then be applied across the growth direction of the sample creating a Schottky diode structure.

**<u>Optical setup:</u>** The sample was cooled to 4K in a closed-loop cryostat to suppress thermal excitation of phonons. Optical excitation and collection were done through a confocal micro-photoluminescence (PL) microscopy setup in the reflection geometry. A combination of half-wave plate and two linear polarizers was used to control the power and polarization of the excitation laser. A schematic of this setup is illustrated in Figure 1c. We used continuous wave (CW) lasers of 532 nm and 880 nm for far-above-band and below-band excitations, respectively. The spot size of the laser is roughly 1.5 µm focused by a 50x NIR long working distance microscope objective. The PL signal was spectrally resolved with a 1200 groove/mm grating in a 50 cm long spectrometer and detected by a Peltier-cooled CCD. This combination produced a spectral resolution of approximately 0.06 nm (~ 0.1 meV).

**PL background subtraction:** A constant dark count background is subtracted from both 532 nm and 880 nm PL data. There is no additional background subtraction for the PL spectra and PL maps obtained under 532 nm excitation. The background in PL spectra obtained with 880 nm excitation do have an energy dependence due to emission from the tail of the wetting layer. This background can be easily fitted by an exponential function and removed to isolate the single dot signals (see Supplemental Information for detail). Backgrounds from both 532 nm and 880 nm excitations are power dependent. Their differences can be seen in Figure 3d of the main text.

**Numerical Simulations:** FDTD simulations of both TE and TM excitations at 532 nm and 880 nm have been conducted by using CST Microwave Studio 2017. CST Microwave Studio is a full-wave 3D electromagnetic field solver based on finite-integral time domain solution technique. A nonuniform mesh was used to improve the accuracy near the Ag slab where the field concentration was significantly large and inhomogeneous. The permittivity data for Ag and GaAs with real dispersion and losses taken from the literature are used.

## Acknowledgements

Work at the University of South Carolina is supported by the ASPIRE internal grant and the National Science Foundation CAREER grant in the Electrical, Communications and Cyber Systems Division (ECCS – 1652720). A.K. and A.A. were supported by the Air Force Office of Scientific Research under award number FA9550-17-1-0002 and by the Welch Foundation with Grant No. F-1802. Sample growth was performed at the Naval Research Laboratory. Y.W. and M.S. would like to thank Professor Yuriy Pershin for the helpful discussions.

## Author Contributions

Y.W. supervised the project and trained M.S on the optical setup. A.B. and D.G. grew the quantum dot sample. M.S. fabricated, characterize the samples and performed all the optical measurements. A.K. conducted the numerical simulations and A.A. supervised the theoretical explorations. M.S. and Y.W. analyzed the data and wrote the paper with input from all authors. All authors discussed the results.

## Supplemental Information

<u>Comparison of PL from empty grooves and silver (Ag) filled grooves</u>

Since the sample in the main text are fabricated with only Ag filled grooves, we fabricated a separate sample from the same wafer which includes empty grooves. The capping layer of this sample was thinned down from 280 nm to roughly 100 nm using the same wet etch solution describe in the Method Section of the main text in an attempt to obtain single dot signals from bare sample. The thinning process produces surface roughness which contribute to the intensity difference seen between the TE and TM polarizations from the bare sample as well as from the empty groove sites. Nevertheless, we can see clearly in Figure S1 that only PLs from the Ag slab exhibit similar drastic polarization dependence presented in the main text.

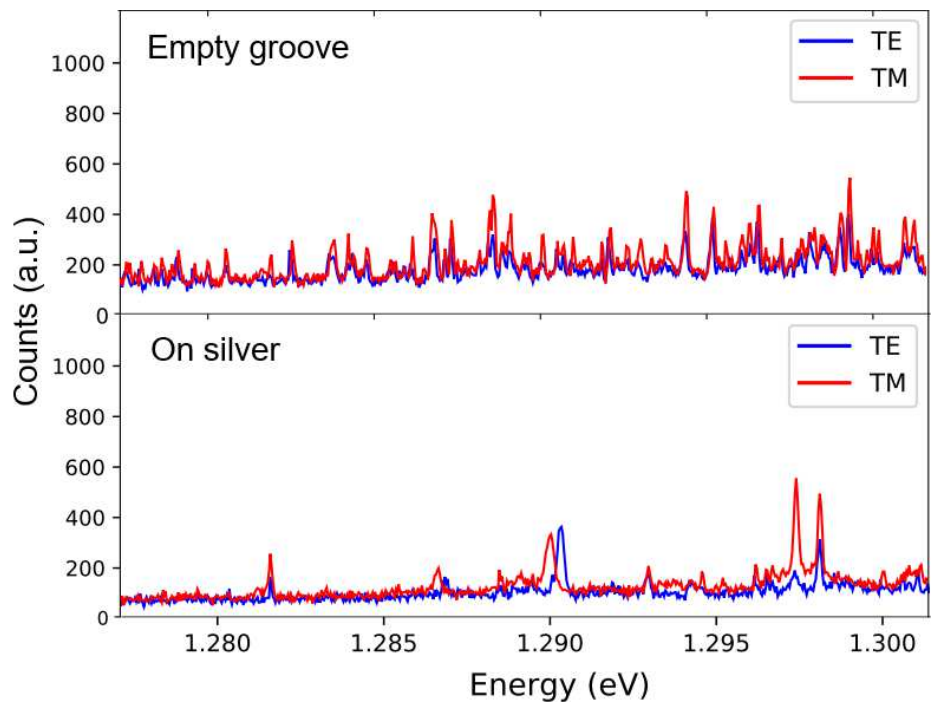

**Figure S1**. Comparison of PL with above-band 543 nm excitation at 15 µW optical power and a bias voltage of 1.45 V with TE and TM excitations on an empty groove and on Ag slab.

Comparison of simulations from straight and rounded groove edges

We performed two sets of simulations using perfect straight groove edges and rounded groove edges according to the lateral AFM profile. In Figure S2, we show that the decreased sharpness of the groove edges leads to noticeably reduced differences between the TE and TM diffracted fields at the plane of the QDs for 532 nm. However, the effect on 880 nm is less due to its longer wavelength.

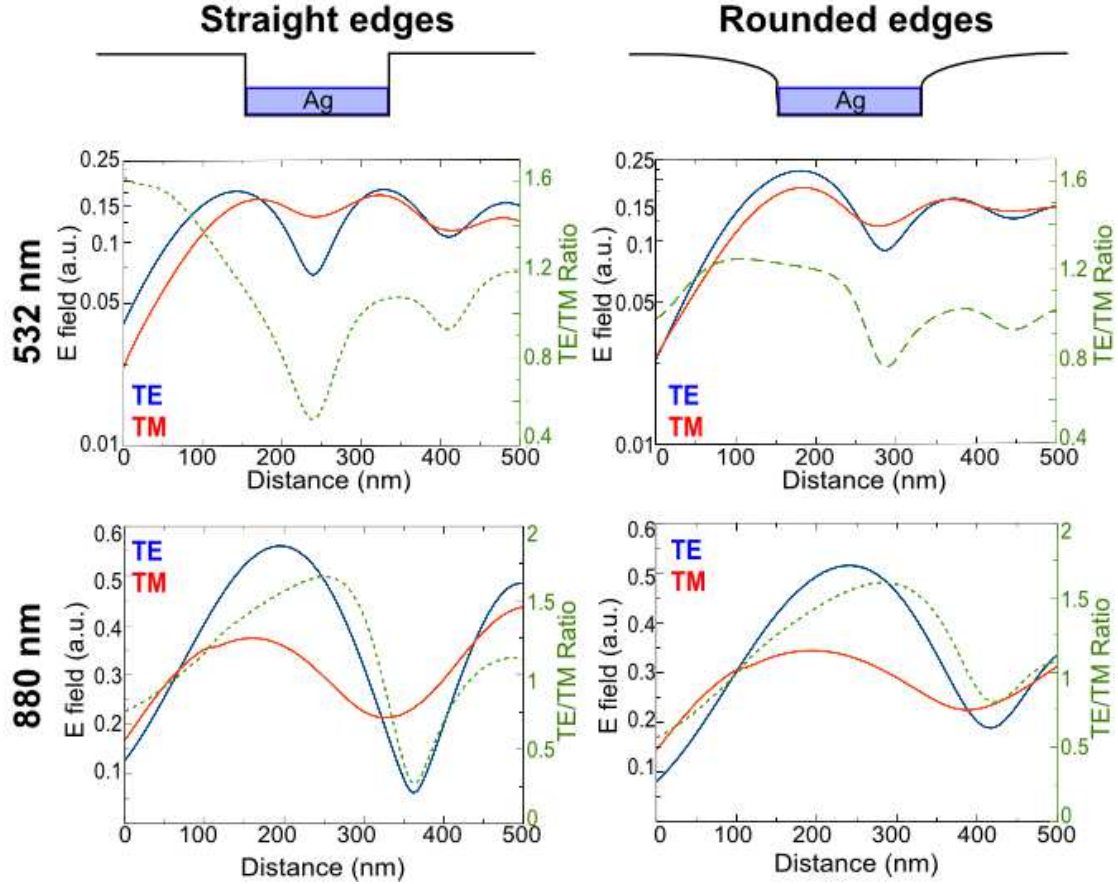

**Figure S2**. Simulations of electric field intensity at the plane of the dots for straight and rounded groove edges for 532 nm and 880 nm illuminations.

Comparison of PL voltage maps with below-band 880 nm excitation

We have performed more detailed PL studies at different sites on and away from the Ag slabs on the same sample presented in the main text. Figure S3 contains a set of representative examples selected from four different excitation sites: two on the bare sample and two on the Ag slabs with TE and TM incident polarizations. The general trend is a reduction in the peak intensity under TM excitation and no apparent change in peak intensity under TE excitation. We have performed a simple statistical analysis on the peak intensity ratio ($R_I = \frac{TE-TM}{TE+TM}$) from PLs obtained from multiple sites on and away from Ag slabs at a bias voltage of 0.5 V. This voltage is chosen because it provides a high peak density in the PL spectra. The results presented in <mark>Figure S4</mark> involve 192 peaks taken from the bare sample and 205 peaks taken on the Ag slab. The average $R_I$ from the bare sample is -0.012 with a standard deviation of $\sigma = 0.056$. These numbers are used as a guide to highlight the differences in the ration $R_I$ calculated from the Ag slab peaks as visualized in <mark>Figure S4</mark>. For the bare sample, 71.9% of the total peaks lies within $\pm\sigma$ and 97.8% lie within $\pm 2\sigma$. In stark contrast, only 38% of the total peaks from the Ag slabs lies within $\pm\sigma$ and 59.5% lie within $\pm 2\sigma$. We plotted $R_I$ as a function of emission energy as well as the maximum intensity count of their respective peaks. We see that the ratios have no notable energy dependence. On the other hand, there is a slight trend of higher $R_I$ for lower counts from the on Ag peaks. It is evident that intensity of TM peaks are statistically lower than that of TE peaks when the excitation is on the Ag slab.

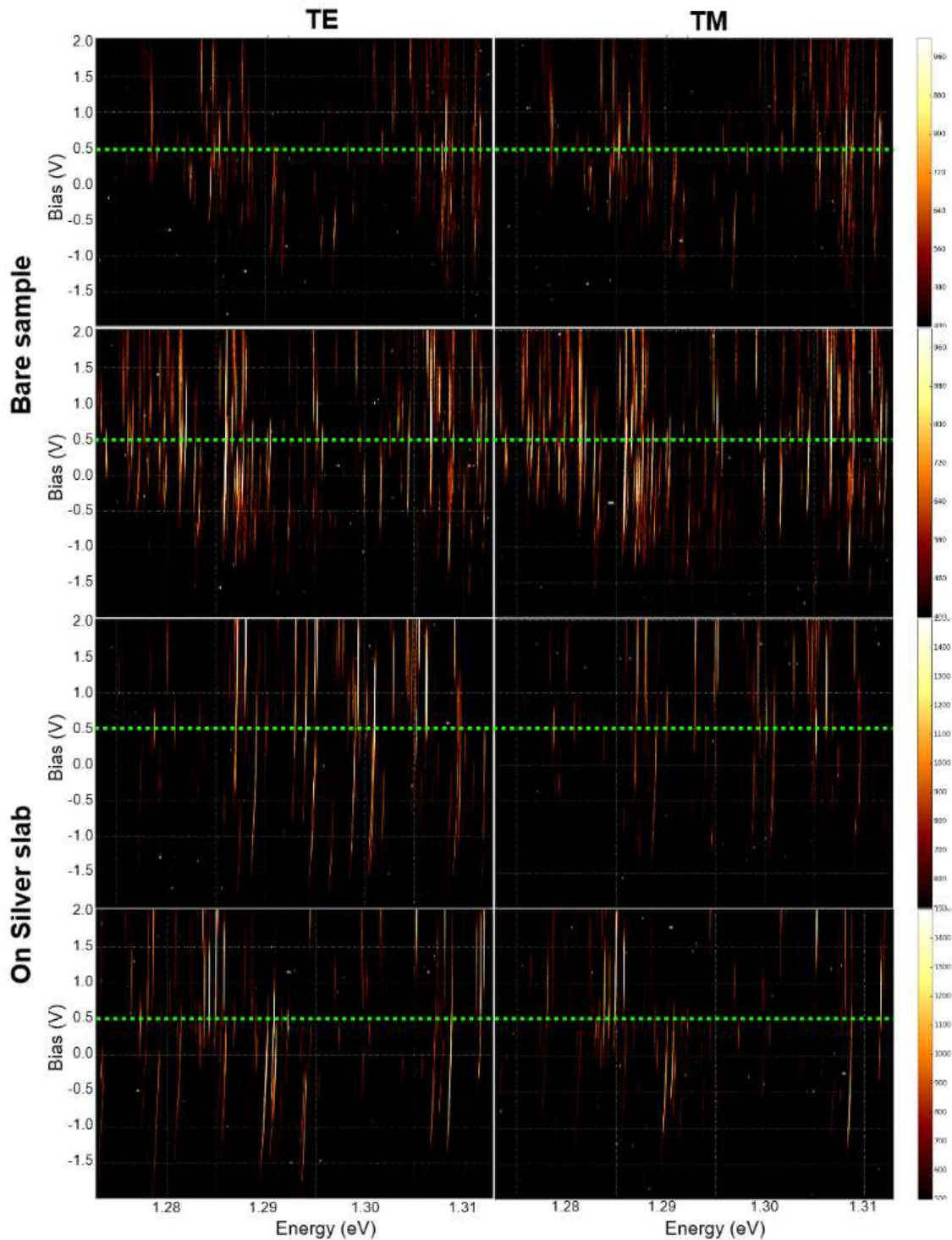

**Figure S3**. Comparison of PL voltage maps with below-band 880 nm excitation at 500 μW optical power. The upper four spectra are taken on bare sample where the lower four spectra are taken from the Ag slab structure. The green dotted lines mark the bias voltage at which data in Figure S4 are taken.

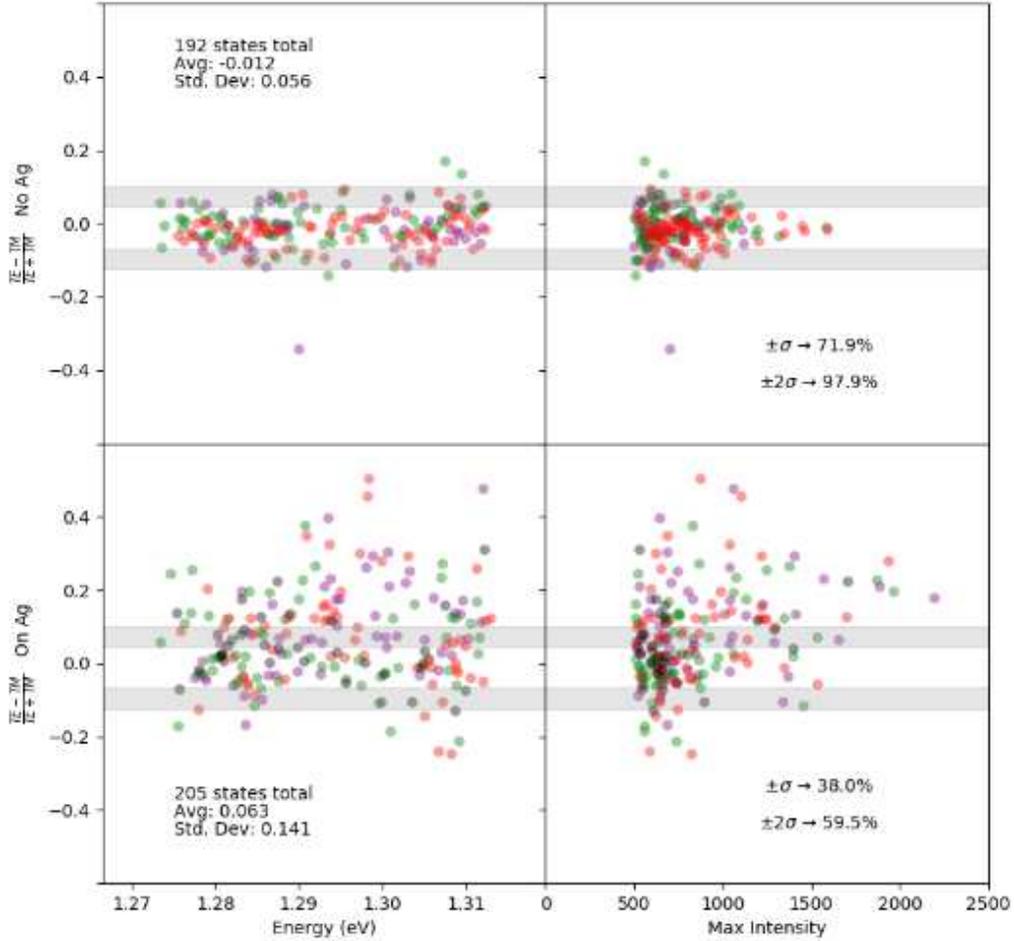

**Figure S4**. The polarization ration $R_I = \frac{TE - TM}{TE + TM}$ as a function of peak energy and peak counts taken with

500 µW of 880 nm excitation at 0.5 V bias. The vertical width of the gray (white) bars are $\pm 2\sigma$ ($\sigma$) wide.

Both color bars are centered at $R_I$ = -0.012 for both on and off Ag plots.

Background subtraction for 880 nm data

The PL spectra and PL maps taken under 880 nm have an energy dependent background due to the tail of the wetting layer. Figure S5 shows a set of TE/TM single PL spectra from 500 µW of

880 nm excitation taken at a bias voltage of 0.5 V before and after background subtraction. We note that the background is exactly the same for TE and TM prior to subtraction. The background is empirically fitted to the exponential function $I(E) = A * e^{B(E-E_0)}$ where $E_0$=1.285 eV and $B = 30.679$. These values are chosen to provide the best fit and are fixed during background fitting for all of the 880 nm spectra. For the 880 nm power dependent background data in Figure 3e, the fit parameter $A$ is extracted and plotted at each excitation power.

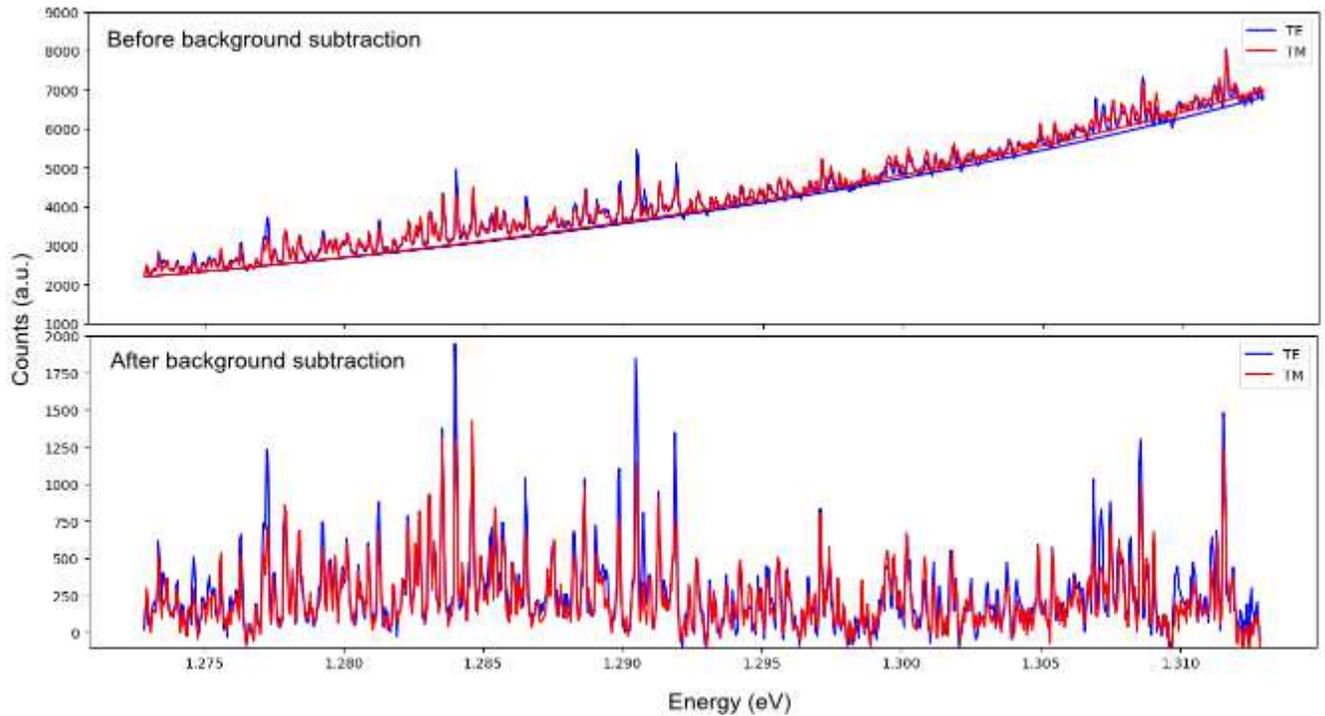

**Figure S5**. PL spectra without and with background subtraction taken with 500 uW of 880 nm excitation at a bias voltage of 0.5 V. The background fits in the top panel are done using an exponential function.